\documentclass[twocolumn,pra,amsmath,amssymb,superscriptaddress]{revtex4}

\usepackage{amsmath, amsthm, amssymb}
\usepackage{graphicx}
\usepackage{dcolumn}
\usepackage{bm}
\usepackage{color}

\def\be{\begin{equation}}
\def\ee{\end{equation}}
\def\tr{\mbox{tr}}

\def\bra#1{\langle#1|} \def\ket#1{|#1\rangle}

\def\proj#1{\ket{#1}\!\bra{#1}}

\newcommand{\ie}{{\it{i.e.~}}}
\newcommand{\etal}{{\it{et al.}}}

\begin{document}

\title{Multipartite fully-nonlocal quantum states}
\author{Mafalda L. Almeida}
\affiliation{ICFO-Institut de Ciencies Fotoniques, E-08860
Castelldefels, Barcelona, Spain}
\affiliation{Centre for Quantum Technologies, National University of Singapore, Singapore}
\author{Daniel Cavalcanti}
\affiliation{ICFO-Institut de Ciencies Fotoniques, E-08860
Castelldefels, Barcelona, Spain}
\affiliation{Centre for Quantum Technologies, National University of Singapore, Singapore}
\author{Valerio Scarani}
\affiliation{Centre for Quantum Technologies, National University of Singapore, Singapore}
\affiliation{Department of
Physics, National University of Singapore, Singapore}
\author{Antonio Ac\'\i n}
\affiliation{ICFO-Institut de Ciencies Fotoniques, E-08860
Castelldefels, Barcelona, Spain} \affiliation{ICREA-Instituci\'o
Catalana de Recerca i Estudis Avan\c{c}ats, Lluis Companys 23,
08010 Barcelona, Spain}


\begin{abstract}
We present a general method to characterize the quantum
correlations obtained after local measurements on multipartite
systems. Sufficient conditions for a quantum system to be
fully-nonlocal according to a given partition as well as being
(genuinely) multipartite fully-nonlocal are derived. These
conditions allow us to identify all completely-connected graph
states as multipartite fully-nonlocal quantum states. Moreover, we
show that this feature can also be observed in mixed states: the
tensor product of five copies of the Smolin state, a biseparable
and bound entangled state, is multipartite fully-nonlocal.
\end{abstract}


\maketitle

\section{Introduction}

Correlations among the results of space-like separated
measurements on composite quantum systems can be incompatible with
a local model~\cite{Bell}. Such phenomenon, known as \emph{quantum
nonlocality}, is an intrinsic quantum feature and lies behind
several applications in quantum information
theory~\cite{nonlocal_prot}. The majority of known results on
quantum nonlocality refer to the bipartite scenario and, even
though multipartite quantum correlations are a potential valuable
resource for multiparty quantum information tasks, their
characterization remains a general unsolved problem.

The most common method to detect nonlocal correlations is through
the violation of a Bell inequality. It is however unclear whether
the amount of violation quantifies nonlocality in a meaningful way
\cite{methot}. For the bipartite scenario, Elitzur, Popescu and
Rohrlich introduced a formalism for the study of nonlocality,
named in what follows EPR-2, which naturally leads to a
quantitative notion of nonlocality \cite{EPR2}. Consider a quantum
state, $\rho$, and a set of local measurements. The obtained joint
probability distribution of outcomes reads $P_\rho(ab|xy)=\tr(\rho
M_x^a\otimes M_y^b)$ where $x$ and $y$ label the measurement
settings, $a$ and $b$ the measurement outcomes, $M_x^a$ and
$M_y^b$ the corresponding measurement operators. The main idea is
to consider all possible decompositions of $P_\rho(ab|xy)$ into a
local and a nonlocal part~\cite{noteloc}:
\begin{equation}\label{EPR2-bipartite}
P_\rho(ab|xy)=p_LP_L(ab|xy)+(1-p_L)P_{NS}(ab|xy).
\end{equation}
The nonlocal distribution $P_{NS}(ab|xy)$ is in principle
arbitrary, but it has to be no-signaling~\cite{notenonsign} since
$P_\rho(ab|xy)$ and $P_L(ab|xy)$ have this property. For the given
state $\rho$, the goal is to identify the decomposition which
maximizes the weight of the local part, $p_L$, among all possible
local measurements. The solution to this optimization problem,
$p_L(\rho)$, is clearly a function of the state only and can be
interpreted as a measure of its nonlocal correlations.

In the bipartite case, some of the most basic questions on quantum
nonlocality have been answered. In what follows it will be useful
to express these findings in terms of the properties of the EPR-2
decomposition~\eqref{EPR2-bipartite}, namely of the weight $p_L$.
For instance, it is known that there exist mixed entangled states
which are local or, equivalently, have
$p_L=1$~\cite{local_entangled}. On the other hand, Gisin's theorem
proves that every pure entangled bipartite state violates a Bell
inequality, which means that they have
$p_L<1$~\cite{gisin_theorem}. For families of two-qubit and
two-qutrit pure states, non-trivial bounds on the value of $p_L$
have been provided~\cite{EPR2, scarani}. The same idea has been
generalized to mixed states in Ref. \cite{zhang}. Moreover, it is
known that fully-nonlocal states exist, as maximally entangled
states have $p_L=0$ \cite{EPR2,BKP}.

Moving to the multipartite scenario, we can find parallel results
to those on bipartite quantum nonlocality. For instance, every
pure entangled multipartite state violates a Bell
inequality~\cite{PR} . Also, no fraction of the statistics
obtained by measuring any state manifesting a GHZ-like paradox can
be described by a completely local
model~\cite{noteloc,GHZparadox,BKP}. However, these results say
nothing about the presence of \emph{genuine multipartite} nonlocal
correlations, \ie those nonlocal correlations established between
all the $m$-parties of a $m$-partite quantum state. In fact, very
few steps were made in the characterization of genuine
multipartite nonlocality. In 1987, Svetlichny provided the first
Bell inequality able to identify genuine tripartite nonlocal
correlations~\cite{svetlichny3}. Much later, this result has been
extended to $m$-partite genuine correlations~\cite{svetlichnyN}.
In Ref.~\cite{bancal} the classification of nonlocal correlations
according to various hybrid local-nonlocal models has been
studied.

In this work, we present a general method to study multipartite
nonlocality, including genuine multiparty, in the no-signaling
scenario~\cite{notenonsign}. It is based on a multipartite version
of the EPR-2 decomposition \eqref{EPR2-bipartite} and on the
results of measurements held on a subset of the parties sharing a
multipartite quantum state. Our framework provides sufficient
conditions to detect genuine multipartite full-nonlocal
correlations, which we simply designate by \emph{multipartite
fully-nonlocal}, in opposition to the (bipartite) full-nonlocality
mentioned previously. We are able to identify the
completely-connected graph states \cite{graph_review} as the first
example of multipartite fully-nonlocal states, proving the
existence of these states for any number of parties. Furthermore,
we prove that multipartite full-nonlocality can also be observed
in the \emph{mixed} case: the tensor product of five copies of the
Smolin state~\cite{Smolin} is a multipartite fully-nonlocal mixed
quantum state.

\section{Characterizing multipartite nonlocality}
In order to introduce the complex structure of correlations in a
multipartite scenario, we start by considering a possible
extension of the EPR-2 decomposition for the tripartite case,
\begin{eqnarray}\label{EPR2-tripartite}
P_\rho(abc|xyz)=p_L P_{L}^{A:B:C}+p_{L:NL}^{A:BC}
P_{L:NL}^{A:BC}+\nonumber\\
p_{L:NL}^{B:AC} P_{L:NL}^{B:AC}+p_{L:NL}^{C:AB} P_{L:NL}^{C:AB}+
p_{NS}P_{NS},
\end{eqnarray}
with $p_L+p_{L:NL}^{A:BC}+p_{L:NL}^{B
:AC}+p_{L:NL}^{C:AB}+p_{NS}=1$. Here, the distribution
$P_L^{A:B:C}$ is completely local \cite{noteloc} and therefore it
strictly contains classical correlations among the outcomes of the
local measurements. On the other hand $P_{L:NL}^{A:BC}$
(equivalently $P_{L:NL}^{B:AC}$ and $P_{L:NL}^{C:AB}$) represents
a local-nonlocal hybrid model
\begin{equation}\label{trihybrid}
P_{L:NL}^{A:BC}=\int d\lambda \omega(\lambda)
P(a|x,\lambda)P(bc|yz,\lambda),
\end{equation}
which was originally introduced by Svetlichny~\cite{svetlichny3}.
We see that measurement results in $A$ are classically correlated
to the results in $B$ and $C$, but nonlocal correlations are
allowed between $B$ and $C$. Note that contrary to what happens in
the bipartite case, the distributions appearing in the different
nonlocal terms of the EPR-2 decomposition~\eqref{EPR2-tripartite}
are arbitrary and may in principle allow signaling between the
corresponding parties. However, here we work in the no-signaling
scenario and, thus, all the terms appearing in the decomposition
are assumed to be compatible with this principle. The intuition is
that the parties cannot signal even if they have access to the
hidden-variable $\lambda$ in \eqref{trihybrid}~\cite{bancal}.
Finally, the component $P_{NS}$ is the only to contain genuine
tripartite nonlocal correlations. This decomposition can easily be
extended to an arbitrary number of parties, $m$. The richness of
multipartite correlations expresses itself by the rapid growth of
hybrid local-nonlocal terms with $m$.

Observe how this multipartite version of the EPR-2 decomposition
clearly distinguishes bipartite from genuine $m$-partite quantum
nonlocality. In order for a quantum state to violate a standard
Bell inequality it is sufficient that it has $p_L<1$, while it
violates a Svetlichny inequality if and only if $p_{NS}>0$.
Analogously, bipartite full-nonlocality is present when $p_L=0$
but multipartite full-nonlocality is synonymous of the much
stronger condition $p_{NS}=1$. Thus, the parameter $p_{NS}$ is the
relevant quantity when studying genuine multipartite nonlocality.
In what follows, we focus our analysis on this quantity and
provide a sufficient criterion to detect multipartite
fully-nonlocal correlations.

For clarity, let us rephrase some known results on multipartite
nonlocality and stress the aim of the present work in terms of the
generalized EPR-2 decomposition \eqref{EPR2-tripartite}. The fact
that every multipartite pure entangled state violates a Bell
inequality~\cite{PR} means that all of them have $p_L<1$. As
commented before, any state exhibiting a GHZ-like paradox has
$p_L=0$~\cite{GHZparadox,BKP}. Here we will provide a sufficient
criterion for a multipartite state to have $p_{NS}=1$. This
criterion identifies several multipartite pure quantum states, as
well as a mixed one, as multipartite fully-nonlocal.

\section{Criteria to detect full-nonlocality in the multipartite scenario}
To start, we introduce a different version of multipartite EPR-2
decomposition \eqref{EPR2-tripartite}, which focuses on the
correlations across a specific bipartition of the composite
system. Consider a $m$-partite state $\rho$ and a bipartition
$\mathcal{A}:\mathcal{B}$, where $\mathcal{A}$ contains $k$
parties and $\mathcal{B}$ the remaining $m-k$. To simplify the
notation, assume that $\mathcal{A}$ contains the first $k$ parties
and $\mathcal{B}$ the $m-k$ remaining ones. Measurement settings
in each partition are labeled by $X=(x_1,\ldots,x_k)$ and
$Y=(x_{k+1},\ldots,x_m)$, and the respective outcomes are
$A=(a_1,\ldots,a_k)$ and $B=(a_{k+1},\ldots,a_m)$. The new version
of the multipartite EPR-2 decomposition is then given by
\begin{multline}
\label{EPR2-multipartite} P_{\rho}(AB|XY)=\\
p^{\mathcal{A}:\mathcal{B}}_{L}P^{\mathcal{A}:
\mathcal{B}}_{L}(AB|XY)+(1-p^{\mathcal{A}:\mathcal{B}}_{L})
P^{\mathcal{A}:\mathcal{B}}_{NS}(AB|XY).
\end{multline}
We use the subscript $L$ to indicate locality in the partition
$\mathcal{A}:\mathcal{B}$, although the distribution
$P^{\mathcal{A}:\mathcal{B}}_{L}(AB|XY)$ is hybrid, \ie
\begin{multline}\label{hybrid local dist}
P^{\mathcal{A}:\mathcal{B}}_L(AB|XY)\equiv \int d\lambda
\omega(\lambda)P(a_1\cdots a_k|x_1\cdots
x_k,\lambda)\\P(a_{k+1}\cdots a_{m}|x_{k+1}\cdots x_{m},\lambda),
\end{multline}
and allows any no-signaling correlations among members of the same
partition. The nonlocal component
$P^{\mathcal{A}:\mathcal{B}}_{NS}(AB|XY)$ in
\eqref{EPR2-multipartite} contains all correlations not modeled by
\eqref{hybrid local dist}. Among all possible decompositions
\eqref{EPR2-multipartite}, we focus on the one maximizing the
weight of the local part, $p^{\mathcal{A}:\mathcal{B}}_{L}$. We
then define full nonlocality with respect to the partition
$\mathcal{A}:\mathcal{B}$ by $p^{\mathcal{A}:\mathcal{B}}_{L}=0$.
It is evident that this multipartite version
\eqref{EPR2-multipartite} strongly resembles the original
bipartite EPR-2 decomposition~\eqref{EPR2-bipartite}: we will see
that this generalized bipartite decomposition form is essential in
the following derivation.

Before showing how to detect multipartite fully-nonlocality, we
must present a method to identify full-nonlocal correlations
across a given bipartition of a $m$-partite quantum state. In
fact, the following theorem constitutes the core of our results.

\textbf{Theorem 1.} An $m$-partite state $\rho$ is fully-nonlocal
across a given bipartition $\mathcal{A}:\mathcal{B}$
($p_L^{\mathcal A:\mathcal B}=0$), in the no-signaling scenario,
if it is possible to create a maximally entangled state between
one party in each partition, for \textit{all} outcomes of suitable
local measurements on the remaining parties~\cite{noteLOCC}.

\emph{Proof:} We are interested in showing that the outcome
distribution of $\rho$, in the EPR-2 decomposition
\eqref{EPR2-multipartite}, has
$p_{L}^{\mathcal{A}:\mathcal{B}}=0$. From \cite{BKP} we know that
any bipartite maximally entangled state is fully nonlocal, \ie,
$p_L=0$. The proof of Theorem 1 will then follow by contradiction:
 $p_{L}^{\mathcal{A}:\mathcal{B}}>0$ would imply $p_{L}>0$ for
the maximally entangled state.

\begin{figure}
\includegraphics[height=0.18\textheight]{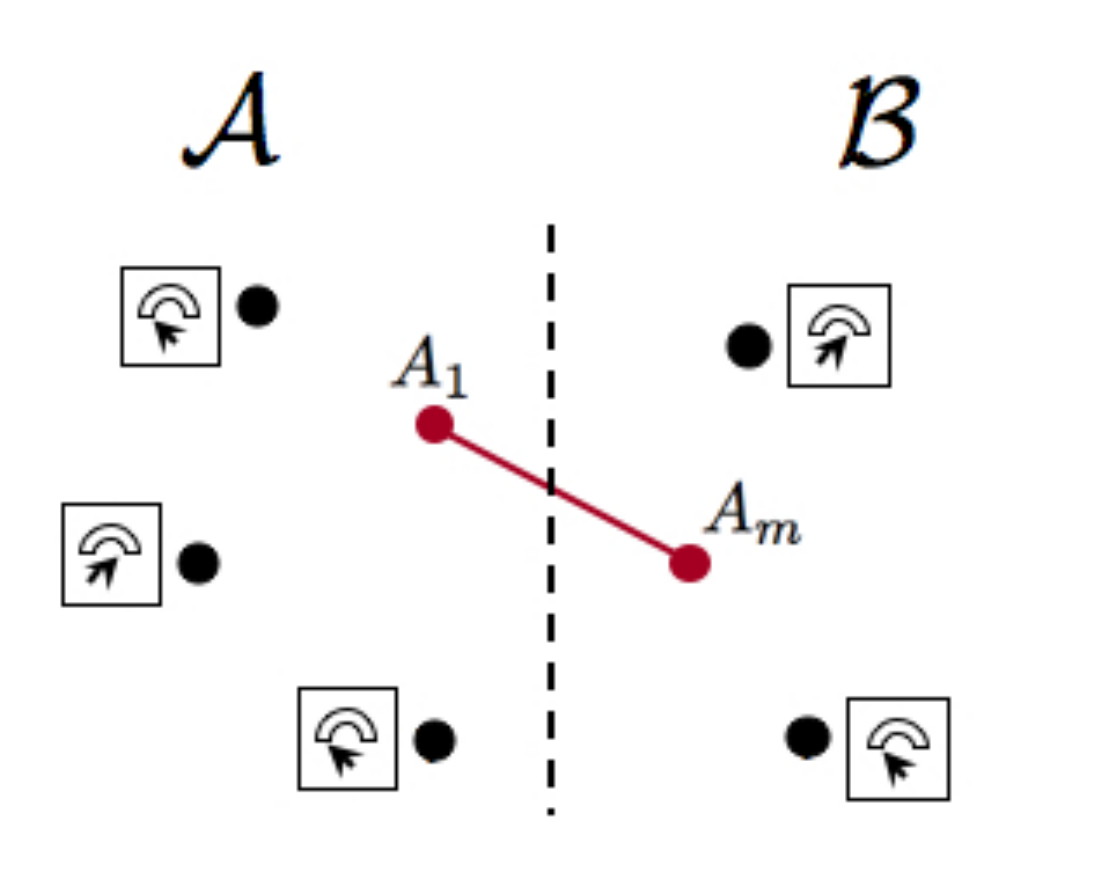}
\caption{(Color online) Local measurements are performed on
parties $A_2\cdots A_{m-1}$ of a $m$-partite state $\rho$.  If for
every outcome of this partial measurement, $A_1$ and $A_m$ share a
maximally entangled state, $\rho$ has $p_L^{\mathcal A:\mathcal
B}=0$ (Theorem 1). If it holds for any possible bipartition, the
state $\rho$ is multipartite fully-nonlocal, $p_{NS}=1$. (Theorem
2)} \label{Fig}
\end{figure}

Indeed, assume there is a positive local weight
$p_{L}^{\mathcal{A}:\mathcal{B}}>0$. For ease of notation,
consider the case in which a maximally entangled state can be
created between party $A_1$, belonging to $\mathcal{A}$, and
$A_m$, belonging to $\mathcal{B}$, by local measurements in the
remaining parties, $M_{2}\otimes \cdots \otimes M_{m-1}$. We label
the measurement settings by $\tilde x$ and $\tilde y$, and their
respective outcomes by $\tilde a$ and $\tilde b$.  For every
outcome $(\tilde a,\tilde b)$, parties $A_1$ and $A_m$ are
projected onto the maximally entangled state $\psi_2^{\tilde
a,\tilde b}$ (See Fig.~\ref{Fig}). Given that we only consider
no-signaling distributions, the following identity holds true for
every component of the EPR-2
decomposition~\eqref{EPR2-multipartite}:
\begin{equation}\label{cond prob} P(a_1a_m,\tilde a\tilde b|x_1x_m,\tilde
x\tilde y)=P(a_1a_m|x_1x_m,\tilde a\tilde b\tilde x\tilde
y).P(\tilde a\tilde b|\tilde x\tilde y).
\end{equation}
This simply expresses the usual no-signaling condition: the
outcome of measurements on parties $A_2\ldots A_{m-1}$ cannot
depend on the choice of measurements by distant parties $A_1$ and
$A_m$.

Since we assumed that $p_{L}^{\mathcal{A}:\mathcal{B}}>0$, a
fraction of the statistics of the state $\rho$ is modeled by the
hybrid model \eqref{hybrid local dist}. Then, for every
measurements choice $(\tilde x,\tilde y)$ on the $N-2$ parties,
there is at least one outcome $(\tilde a,\tilde b)$ which the
hybrid model predicts with non-zero probability: $P_L^{\mathcal
A:\mathcal B}(\tilde a\tilde b|\tilde x \tilde y)>0$. The
post-measurement state associated to this outcome, $\psi_2^{\tilde
a,\tilde b}$, is maximally entangled by assumption and has
correlations according to the induced EPR-2 decomposition
\begin{multline}\label{induced decomp mes}
P_{\psi_2}(a_1a_m|x_1x_m)=
p_LP_L^{\mathcal{A:B}}(a_1a_m|x_1x_m,\tilde a \tilde b\tilde x
\tilde y)+ \\
(1-p_L)P_{NS}^{\mathcal{A:B}}(a_1 a_m|x_1x_m,\tilde a \tilde b
\tilde x \tilde y) ,
\end{multline}
with local weight
\begin{equation}\label{local ind mes}
p_L=\frac{p_L^{\mathcal A:\mathcal B}}{P_L^{\mathcal{A:B}}(\tilde
a \tilde b |\tilde x \tilde y)}{P_{\rho}(\tilde a \tilde b |\tilde
x \tilde y)}.
\end{equation}
Notice that the induced nonlocal distribution is well-defined
because its only constraint, the no-signaling condition, is not
affected \cite{noteNSdist}. Also, the fact that we can apply the
no-signaling condition to every component of the hybrid model
\eqref{hybrid local dist}, guarantees that we obtain the valid
induced hybrid distribution
\begin{multline}
P_L(a_1b_1|x_1y_1,\tilde a \tilde b \tilde x \tilde y)= \\ \int
d\lambda \tilde\omega(\lambda) P(a_1|x_1,\tilde a \tilde x
\lambda)P(b_1|y_1,\tilde b \tilde y \lambda).
\end{multline}
where the induced probability density is
\begin{equation}
\tilde\omega(\lambda)=\frac{\omega(\lambda)P(\tilde a|\tilde
x,\lambda)P(\tilde b|\tilde y,\lambda)}{P_L(\tilde a\tilde
b|\tilde x\tilde y)}.
\end{equation}
According to \eqref{local ind mes}, the local weight $p_L$ is
necessarily positive. This is known to be impossible since it
corresponds to a maximally entangled state $\psi_2^{\tilde
a,\tilde b}$~\cite{BKP}. Therefore, the distribution for the state
$\rho$ must be fully nonlocal on the partition $\mathcal
A:\mathcal B$, $p_L^{\mathcal A:\mathcal B}=0$. $\square$

An important remark on the previous result is that the proof is
presented as a sequence of measurements only by clarity reasons.
In fact, given that all  measurements are performed on
spatially-separated parties, the results of measurements on
parties $A_2\ldots A_{m-1}$ are guaranteed to be independent from
the measurement choices of parties $A_1$ and $A_m$ by the
no-signaling principle. Therefore, there is no need to impose any
time-ordering on the measuring events and we are in the most
standard framework of nonlocality, which considers single local
measurements in space-like separated systems.

We are now ready to finally present the result which, combined
with Theorem 1, provides the sufficient criterion to detect
multipartite full-nonlocality.

\textbf{Theorem 2.} A probability distribution is multipartite
fully-nonlocal ($p_{NS}=1$) if and only if it is fully nonlocal
($p_L^{\mathcal{A}:\mathcal{B}}=0$) in every bipartition
$\mathcal{A}:\mathcal{B}$.

\emph{Proof:} The proof proceeds again by contradiction. Assume
that $p_{NS}<1$. Then, there is at least one local/hybrid model in
the EPR-2 decomposition \eqref{EPR2-tripartite} with positive
weight. However, there is always a bipartite splitting of the
parties such that this term contributes to the corresponding local
part in Eq.~\eqref{EPR2-multipartite}. But this is now in
contradiction with the fact that $p_L^{\mathcal{A}:\mathcal{B}}=0$
for every bipartition. Then, $p_{NS}$ is equal to one for the
initial distribution.$\square$

\section{Multipartite fully-nonlocal states}

\subsection{Completely-connected graph states}

From Theorems 1 and 2 we can immediately identify the
completely-connected graph states as being multipartite
fully-nonlocal states. This comes from the fact that these states
fulfill all the necessary requirements: for any pair of qubits,
there are local Pauli measurements on the remaining $N-2$ parties
which project the pair of particles into a maximally entangled
state for every measurement outcome~\cite{graph_review,graph
states}.

Graph states are known to possess several peculiar features like
being perfect quantum channels for quantum communication, or (some
of them) universal resources for measurement-based quantum
computation \cite{graph_review}. The fact that
completely-connected graph states are multipartite fully-nonlocal
is one more interesting feature of this important class of
multipartite entangled states.

\subsection{A multipartite fully-nonlocal mixed state}

We now present the first known example of mixed state which
contains multipartite fully-nonlocality. This example is based on
the so-called Smolin state, a four-partite bound entangled state
given by \cite{Smolin}
\begin{equation}
\rho^S_{ABCD}=\frac{1}{4}\sum_{i=0}^3\proj{\psi_i}_{AB}\otimes\proj{\psi_i}_{CD},
\end{equation}
where $\ket{\psi_i}$ denote the four Bell states.

The Smolin state is biseparable among all two-qubit versus
two-qubit partitions, and consequently no entanglement can be
distilled from it by any local operations and classical
communication. Interestingly, it was shown in
Ref.~\cite{SupraActivation} that the combination of five of these
states, namely
\begin{equation}
M^S=\rho^S_{ABCD}\otimes\rho^S_{ABCE}\otimes\rho^S_{ABDE}\otimes\rho^S_{ACDE}\otimes\rho^S_{BCDE}\,.
\end{equation}
is distillable. Actually, Ref.~\cite{SupraActivation} presents a
distillation protocol which deterministically transforms $M^S$
into a singlet state between any pair of parties. We can then
apply our sufficient criterion to conclude that $M^S$ has
$p_{NS}=1$. This result proves the existence of mixed states which
contain multipartite fully-nonlocal quantum correlations.


\section{Conclusion} We have seen how generalizations of the EPR-2 decomposition for quantum probability distributions and outcomes
of partial measurements on quantum systems can be used to study
multipartite nonlocal correlations. Our formalism  gives
sufficient conditions to detect multipartite full-nonlocality and
identifies all completely-connected graph states as examples of
multipartite fully-nonlocal states. This result solves a
fundamental question concerning the characterization of nonlocal
quantum correlations: in the no-signalling scenario, multipartite
fully-nonlocal states exist for any number of parties. In
addition, we also provide an example of such extreme nonlocality
for a multipartite mixed state.

Finally, our work opens new questions on the characterization of
multipartite nonlocality. Would our conclusions be affected if the
different terms appearing in the decomposition were not
constrained by the no-signaling principle? Providing this extra
resource could give the hybrid models the ability to reproduce a
fraction of the quantum probability distribution, and then
full-nonlocality would be lost. Another interesting open question
is to characterize the set of all multipartite fully-nonlocal
quantum states. Is the derived sufficient criterion also necessary
to identify multipartite fully-nonlocal quantum correlations?

\begin{acknowledgements}
The authors thank  S. Pironio and M. Grassl for helpful
discussions. This work was financially supported by the
Funda\c{c}\~{a}o para a Ci\^{e}ncia e a Tecnologia (Portugal)
through the grant SFRH/BD/21915/2005, the European QAP and PERCENT
projects, the Spanish MEC FIS2007-60182 and Consolider-Ingenio
QOIT projects, Generalitat de Catalunya, Caixa Manresa and the
National Research Foundation and the Ministry of Education of
Singapore.
\end{acknowledgements}

\bibliographystyle{plain}

\begin{thebibliography}{99}
\bibitem{noteloc}
A joint probability distribution is local whenever it can be
written as $P(ab|xy)=\int d\lambda \omega(\lambda)
P(a|x,\lambda)P(b|y,\lambda)$ where the classical variable
$\lambda$ is distributed according to the probability distribution
$\omega(\lambda)$. This definition easily generalizes to an
arbitrary number of parties. Those probability distributions which
do not admit this decomposition are said to be nonlocal.
\bibitem{notenonsign}
A distribution $P(ab|xy)$ is no-signaling if the local
distributions for one party do not depend on the choice of
measurements by the other party, for instance $P(a|xy)\equiv
\sum_{b}P(ab|xy)=P(a|x)$. This definition easily generalizes to an
arbitrary number of parties.
\bibitem{noteLOCC}
Actually, our proof also applies for slightly more general
protocols in which the local measurements among parties belonging
to the same block in the bipartition can be correlated using
classical communication. It also applies for situations where parties $A_1$ and $A_m$ are projected into any fully nonlocal state ($p_L=0$).
\bibitem{noteNSdist} If $P_{NS}^{\mathcal A,\mathcal B}(\tilde a\tilde b|\tilde x,\tilde
y)=0$, the nonlocal distribution is not well-defined but the
induced state $\psi_2^{\tilde a \tilde b}$ is local and our proof
still holds.
\end{thebibliography}

\end{document}